\documentclass[12pt]{iopart}
\usepackage{CJK}
\usepackage[style = ieee, sorting=none, url=false, isbn=false]{biblatex}
\usepackage{biblatex}
\AtEveryBibitem{\clearfield{month}}
\AtEveryBibitem{\clearfield{day}}
\usepackage{graphicx} 
\usepackage{caption}
\usepackage{hyperref}
\usepackage{xcolor} 
\usepackage{amsmath} 
\usepackage{amssymb} 
\usepackage[percent]{overpic} 
\usepackage{subcaption}  
\usepackage[normalem]{ulem} 
\usepackage{multirow} 

\newcommand{\vect}[1]{\boldsymbol{#1}} 

\DeclareMathOperator{\arsinh}{arsinh}

\begin{document}
\title[Local flow measurements around flexible filaments under
rotating magnetic field]{Local flow measurements around flexible filaments under
rotating magnetic field}

\author{Andris P\={a}vils Stikuts, Abdelqader Zaben, Ivars Dri\c{k}is, M\={a}ra \v{S}mite, R\={u}dolfs Livanovi\v{c}s, Andrejs C\={e}bers, Guntars Kitenbergs}

\address{MMML lab, Department of Physics, University of Latvia, Riga, Jelgavas 3, LV1004, Latvia}
\ead{guntars.kitenbergs@lu.lv}
\vspace{10pt}

\begin{abstract}
Effective mixing of fluids at the microfluidic scale is important for future applications in biology, medicine, and chemistry.
A promising type of micromixers are magnetic filaments, which can be activated by an external magnetic field.
However, there is a lack of research that combines experiments and numerical modelling of hydrodynamics around such filaments.
Here we use micro-particle image velocimetry to measure flow fields around rotating flexible ferromagnetic filaments and compare them to numerical data from elastic rod model.
We measure that rotating filaments hover above the surface, whereas the resulting fluid velocities are highly dependent on the hovering distance.
We also find that the rotating filament causes a 3D flow coming from the rotational plane and being extracted along the axis of rotation.
These findings will help develop better micromixers.
\end{abstract}

%
%
%
%
%

\section{Introduction}

 Mixing in a low Reynolds number regime is an increasingly active topic of research because of its importance in microfluidics and development of lab-on-a-chip devices. 
 Mixing is required for various applications, particularly in the fields of biology, medicine, and chemistry \cite{LEE2018677}.
 A promising approach is magnetic mixing or magnetic stirring, where mixing is done with magnetic materials and an external magnetic field \cite{cao_recent_2020}.
 Although it is an active mixing method, application of an external field is easy to implement, and there is a wide selection of magnetic materials.
 An interesting active micromixer that shows promising efficiencies as presented by D.Owen \textit{et al.} \cite{OWEN201684}, is realised by means of actuating magnetic beads under a rotating magnetic field. 
 Here, we focus on micrometre-sized ferromagnetic beads linked to form a flexible filament and driven by an external rotating magnetic field for stirring. 
 Their advantages include commercial availability and relatively low cost for filament synthesis \cite{mi10110731}. 
 Besides the on-demand control of mixing, magnetic filaments have been shown to be a promising candidate for cargo delivery \cite{PhysRevE.79.051503,20}, sensing \cite{B918215B}, and microrheology \cite{C4SM02454K}.       
  
  The dynamics of flexible ferromagnetic filaments under the action of a rotating magnetic field has been investigated by our group both numerically \cite{PhysRevE.96.062612} and experimentally \cite{zaben_deformation_2020-1,zaben_3d_2020}.
  These studies allow us to estimate the physical properties and deformation of the filament shape under the action of rotating magnetic field. 
  Moreover, it was found that ferromagnetic filaments take a 3D shape when operating at higher frequencies \cite{zaben_3d_2020}.
  This is different for flexible paramagnetic filaments where back-and-forth motion is observed for higher frequencies \cite{biswal_rotational_2004}.
  This also limits the mixing rate for paramagnetic chains, where a rapid drop in efficiency occurs after an optimal frequency \cite{Biswal}. 
  The difference in behaviour gives us the motivation to further investigate the use of ferromagnetic filaments as mixers.  
 
  For a better understanding of the physics related to micromixing, it is esential to characterise the hydrodynamics around such filaments.
  Experimentally, this includes measurements of flow fields, accessible by the micro Particle Image Velocimetry ($\mu$PIV) technique.
  Although challenging due to resolution limits, this approach has been used to study hydrodynamics at low Reynolds number for artificial micron sized swimmers and microorganisms, including L-shaped swimmer, driven by a rotating magnetic field \cite{mi10120865}, flow measurements around microalgae \cite{PhysRevLett.105.168102} and 3D flow measurements for magnetically actuated artificial cilia using stereoscopic $\mu$PIV \cite{chen_pekkan_2013}.
  However, coupling of experimental measurements around a simpler model system flexible filament with the corresponding numerical simulations has not been done.
  We aim to solve this problem with this paper.
  First, the experimental system is described, including the steps to improve the resolution.
  Second, the numerical model is introduced, and the important aspect of the filament's height above the surface is discussed.
  Finally, we compare the experimental and numerical results, noticing and characterising a 3D flow, induced by the filament rotation.

\section{Experimental system}

\subsection{Filament Synthesis}
  For filament synthesis we follow the method proposed by Dreyfus \textit{et al.} \cite{dreyfus_microscopic_2005}.
  It uses the formation of a streptavidin-biotin bond between DNA fragments and magnetic particles, aligned by a static magnetic field. 
  In detail, this method has been described by K.\={E}rglis \cite{erglis_experimental_2010}, and was further adopted for our earlier studies on filament dynamics under a rotating field \cite{zaben_deformation_2020-1,zaben_3d_2020}. 
  The ferromagnetic particles (Spherotech, $1~\%$w/v) are made of polystyrene and are covered with a layer of chromium oxide, functionalised with streptavidin, and have an average diameter of 4.26~$\mu$m. DNA fragments (Latvian Biomedical Research and Study Centre), biotinylated at the 5' ends, have a length of 1000bp and a concentration of $192~\mu$g/ml. 
  
  Samples are made with the following procedure. 40$\mu$l of $0.53~\mu$m tracer Nile red fluorescent particles (Spherotech, $1~\%$w/v), for particle displacement measurements, is mixed with $0.5$~ml of $10\% $ TE buffer solution ($\eta = 0.001$~Pa, pH = 7.5), resulting in a concentration of $0.08~\%$(w/v). 
  $10~\mu $l of 0.01\% Sodium Dodecyl Sulphate (SDS) is then added and the sample is sonicated for 30~min, to reduce particle aggregations. 
  $10$~$\mu$l DNA (6.2~nM) and $2$~$\mu$l of magnetic particles ($0.004\%~w/v$) are then added to the sample, and placed for two minutes between two Neodymium magnets, creating a homogeneous field of $\approx 50$~mT.
   
  \subsection{Experimental setup}
    
  The experimental setup consists of a micro Particle Image Velocimetry ($\mu$PIV) system combined with a coil system for magnetic field generation.
  
  The $\mu$PIV setup (Dantec Dynamics) consists of an inverted optical microscope (Leica DM-ILM with a Y3 filter cube, 40x objective), a HiSense MkII camera for image acquisition, which has a CCD sensor with double frame mode (maximum frame rate $6.1$~Hz, minimum interframe time $200$~ns, double pulsed YAG laser ($\lambda = 532$~nm) which is used as a light source for exciting the florescent particles and has a maximum frequency of $50$~Hz. 
  The camera and the laser are synchronised through a timer box (NI PCI-6602) and controlled by DynamicStudio software.

  Magnetic field generation is done by an in-house built coil system, which has 3 pairs of coils to generate magnetic field in the three dimensions. 
  The coils are connected to 3 AC power supplies (Kepco BOP 20-10M), giving a maximum current of $7$~A, which correspond to fields up to $12$~mT. 
  The desired field profile is generated by a Labview code.
  For a rotating field profile, two sine signals with a phase shift of 90 degrees are generated and sent through National Instruments data acquisition card (NI PCI-6229) that is connected to the power supplies.

  \subsection{Experimental procedure}

For observation we use fluidic cells, which are made from two glass slides separated by a $211$~$\mu$m thick double-sided adhesive tape.
Experiments were conducted as follows.
$20$~$\mu$l of the prepared filament sample is pipetted in the fluidic cell and placed under the microscope. 
The acquisition mode is set to double frame mode while the inter-frame time and the acquisition frequency is chosen based on the field frequency.
These parameters are given later. 
For each experiment, double frames are acquired for filaments that complete $9$ to $25$ rotations. 
The laser power was set to $100\%$ and a maximum attenuator level of $25\%$. 
For higher levels, the filament was found to break.
Due to gravity, the filaments were found to sediment to the bottom of the fluidic cell, with free particles around, which may have detached during the transfer process.
A rotating field ($B=1.72$~mT, $f=1$~Hz) is applied first for 10 to 15 minutes to allow the free particles to connect to the filament. 
During the rotations, filaments were also found to have a slight shift, likely due to interactions with the wall, which differs for particles with different sizes.

\begin{figure}[h]
\centering
 \includegraphics{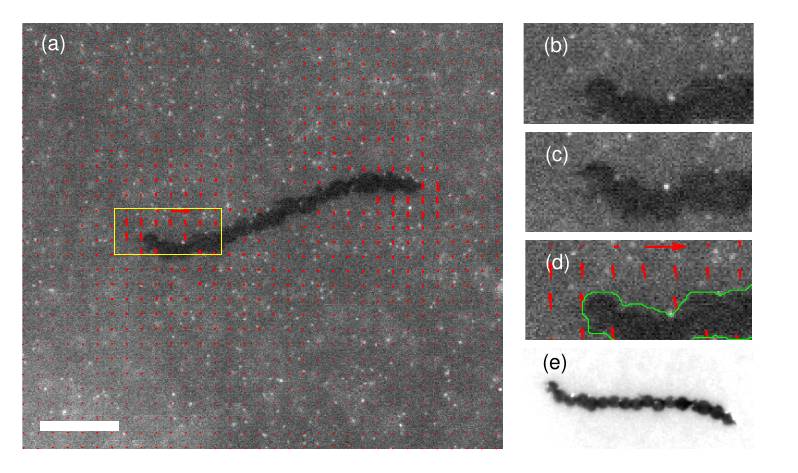}
  \caption{An example of experimental image and image processing. (a) Image of a rotating filament (filament 0 at $f=3$~Hz \& $B=3.44$~mT, clockwise direction) within tracer particles with a superimposed velocity field (red arrows). Yellow recatngle indicates zoom-in region. Scale bar is $20~\mu$m. Zoom-in region of the first (b) and the second (c) frame of the double frame image. (d) Zoom-in region with velocity field and border of filament mask (green). (e) Superimposed image of all filaments, after the correction for the center of rotation.}
  \label{fig:exp_proc}
\end{figure}

An example of an acquired image is given in figure~\ref{fig:exp_proc}, where (a) shows the rotating filament with fluorescent tracer particles around, while (b) and (c) visualise the first and second frames of the double frame mode on a small zoom-in region.
For better visability, the contrast of these images is increased.

\subsection{Particle Image Velocimetry and data processing}

 For processing, we export the experimental images from Dynamic Studio and do initial image processing in Matlab. 
 We first mask the filament to eliminate its signal in the cross-correlation calculations, and we detect only the flow profile around the filament. 
 The filament was first segmented, based on intensity thresholding, then dilated to increase the mask size to remove any tracer particles that may got attached to the filament during the experiment.
 An example of the masked region is indicated in figure~\ref{fig:exp_proc}~(d) with a green border. 

 Patricle image velocimetry (PIV) analysis was performed using PIVlab, an open source software developed by W. Thielicke and E.J. Stamhuis \cite{PIVlab}. 
 The PIV algorithm used here is based on Fourier transform cross-correlation, and the analysis was made by defining the interrogation area as $48\times48$~pixels with a PIV step size of $16$ pixels. 

As the PIV measurement takes place close to the resolution limit, the resulting velocity values are noisy.
An example of flow field obtained from a single double frame image is given by the red vectors in figure~\ref{fig:exp_proc}~(a)\&(d).
A qulitative character of the flow field can be obtained, but the pecularities are lost in noise. 
Such noise is usually eliminated by time averaging.
In our experiments, the filament rotates with its centre of rotation moving, so summing the velocity fields simply frame by frame is not an option. 
We use Python for further experimental image and data processing and solve this problem as follows.

The first step is a geometric preprocessing. 
We find the filament's rotation angle in each frame, by selecting the data sequence where the rotation rate is steady.
Due to the chosen camera framerate and field frequency, there are only a small number (1, 2, 3 or 6) of filament rotation angles against a fixed direction.. 
We find the precise angle values by cross-correlating images generated by the forward Radon transform, obtained using functions from scikit-image library \cite{van_der_walt_scikit-image_2014}.
To find the filament displacements during the experiment, we first find the displacements for filaments with the same angle of rotation by cross-correlation. 
As a result, filaments with the same angle of rotation are aligned with each other.
Finally, a common centre of mass is found by the least-squares method. 
The quality of this can be assessed in figure~~\ref{fig:exp_proc}~(e), which shows a superimposed image of all filament images for the particular filament, field and frequency. 
Once the offsets of the centre are found, we calculate the coordinates for the centre of rotation for each of the frames above.

\begin{figure}[h]
\centering
 \includegraphics[width=\textwidth]{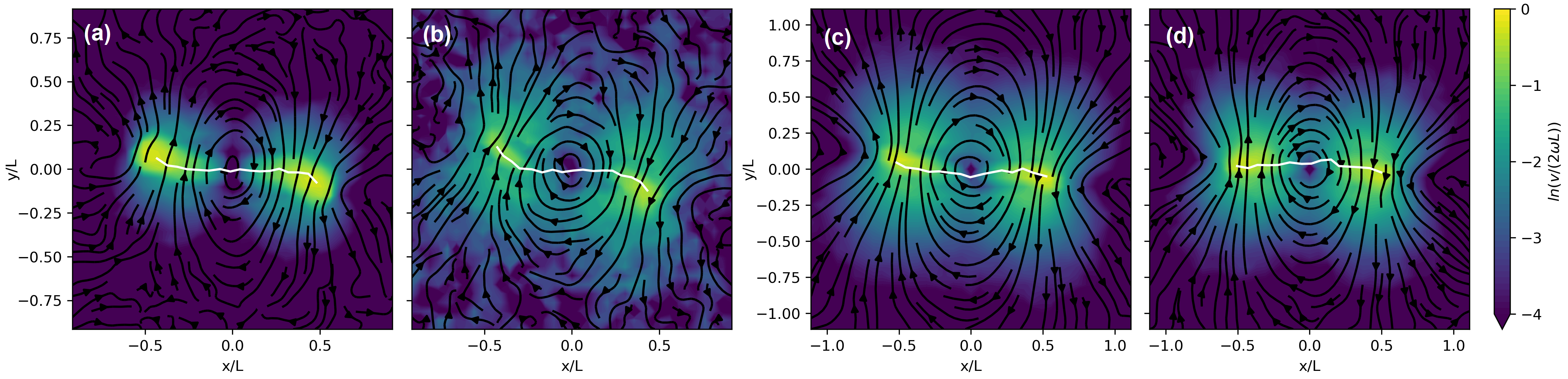}
  \caption{Examples of experimentally measured and averaged flowfields including streamlines and normalized velocity amplitudes around several filaments. Filament 0 at (a) $f=1$~Hz and (b) $f=5$~Hz, and filament 1 (c) and filament 3 (d) at $f=3$~Hz. See table~\ref{tab:sim_params} for more details. Each graph is $130\times130~\mu$m$^2$.}
  \label{fig:exp}
\end{figure}

In the second step, we average the velocity fields.
First, we shift and rotate each PIV velocity field so that the centre of rotation is at the origin and filament's centre line is parallel to the $x$ axis. 
It is important to note that the grids of points where the velocities are determined by the PIV do not coincide.
To overcome this, we create a new grid of points where we want to find the velocity field values with a chosen gridsize, here fixed at $16$~px.
For each of the grid points we calculate the velocity values as the average of the PIV velocity measurement points that fall within a chosen radius around this point, here chosen as $9$~px.
The same approach is used when calculating velocities along other paths, e.g. lines or circles, used for characterising flows.
Several examples of experimentally measured and averaged flow fields are given in figure~\ref{fig:exp}, while the averaged flowfield corresponding to the figure~\ref{fig:exp_proc}~(a) is given in the middle bottom graph of figure~\ref{fig:exp_sim_velocity}. 
 
\section{Numerical model}
\label{sec:numerical_model}
The filament is numerically modelled using the Kirchhoff theory of elastic rods \cite{landau_theory_2009, powers_dynamics_2010}, where the effect of the magnetic field is added \cite{doi:10.1002/adfm.201502696}.
The filament is described by its centre line $\vect y(l)$, which is parameterized by its arc length $l\in[-L/2,L/2]$, where $L$ is the length of the filament.
The cross-section of the filament experiences a force
\begin{equation}
    \vect F = -A \frac{\partial^3\vect y}{\partial l^3} - \mu_0 M \vect H + \Lambda \frac{\partial \vect y}{\partial l},
\label{eq:filament_force}
\end{equation}
where $A$ is the bending modulus, $\mu_0$ is the vacuum permeability, $\vect H$ is the external magnetic field intensity and $M$ is the filament's linear magnetization (magnetic moment per unit length), and $\Lambda$ is the tension force to ensure the inextensibility of the filament.
The velocity of the filament is determined using the resistive force theory, which states that the velocity is proportional to the linear force density
\begin{equation}
    \vect v = \frac{\vect f_{\parallel}}{\zeta_\parallel} + \frac{\vect f_{\perp}}{\zeta_\perp},
\end{equation}
where $\vect f = \partial \vect F / \partial l$ is the linear force density, $\parallel$ denotes the projection along the tangent of the filament, $\perp$ denotes the projection normal to the filament, and $\zeta_\parallel$ and $\zeta_\perp$ are the drag coefficients.
We enforce free end boundary conditions (zero force and torque at $l=\pm L/2$), which give $\vect F|_{l=\pm L/2}=\vect0$ and $ \partial^2 \vect y / \partial l^2 |_{l=\pm L/2}=\vect0$.

The drag coefficients are affected by the presence of the no-slip wall below the filament. 
For a filament oriented and moving parallel to the wall, the drag coefficients can be approximated as \cite{koens_local_2021}
\begin{equation}
\label{eq:drag_coefs}
\begin{split}
    \zeta_\parallel &= \frac{4\pi\eta}{
    \ln{\left( \frac{L^2}{4 h^2} \right)}-1-E_1+E_2+E_3+2\alpha
    },\\
    \zeta_\perp &= \frac{8\pi\eta}{
    \ln{\left( \frac{L^2}{4 h^2} \right)}+1-E_1-2E_2+2\alpha
    },
\end{split}
\end{equation}
where $\eta$ is the viscosity of the surrounding fluid and
\begin{equation}
\begin{split}
    E_1 &= 2 \arsinh{\frac{L}{4 h}}, \\
    E_2 &= \frac{1}{2\sqrt{1+16\frac{h^2}{L^2}}}, \\
    E_3 &= \frac{1}{2( 1+16\frac{h^2}{L^2})^{3/2}}, \\
    \alpha &= \ln{\left(   \frac{h + \sqrt{h^2-a^2}}{a} \right) },
\end{split}
\end{equation}
where $\arsinh$ is the inverse of the hyperbolic sine, $h$ is the height above the bottom wall measured from the centerline of the filament, and $a$ is the filament's cross-section radius.
Far away from the wall $h/L\rightarrow\infty$, the drag coefficients take the familiar \cite{blake_note_2010} values of
\begin{equation}
\begin{split}
    \zeta_\parallel = \frac{2\pi\eta}{
    \ln{\left( \frac{L}{a} \right)} - \frac{1}{2}
    },\quad
    \zeta_\perp = \frac{4\pi\eta}{
    \ln{\left( \frac{L}{a} \right)} + \frac{1}{2}
    }.
\end{split}
\end{equation}
However, they diverge to infinity as $h\rightarrow a$.

The filament is discretized into N nearly cylindrical segments, and the velocity of each segment is determined by 
\begin{equation} 
\label{eq:velocity_calculation}
\begin{split}
    \vect v_1 &= \frac{1}{\Delta l}\left( \frac{\vect F_{1\parallel}}{\zeta_\parallel} + \frac{\vect F_{1\perp}}{\zeta_\perp} \right), \\
    \vect v_i &= \frac{\vect f_{i\parallel}}{\zeta_\parallel} + \frac{\vect f_{i\perp}}{\zeta_\perp},\quad i=2\ldots N-1 \\
    \vect v_N &=  -\frac{1}{\Delta l}\left( \frac{\vect F_{N\parallel}}{\zeta_\parallel} + \frac{\vect F_{N\perp}}{\zeta_\perp} \right),
\end{split}
\end{equation}
where the numerical index denotes the segment number, $\Delta l$ is the length of the cylindrical segment.
Since the magnetic term in eq. \eqref{eq:filament_force} is constant, only the first and the last segment directly experience the magnetic force.
The filament is assumed to be inextensible; therefore, the calculated velocities are projected on the space of inextensible motion using the procedure described in ref. \cite{nedelec_collective_2007}.
This permits us to bypass the direct tension force calculation.
The derivatives with respect to $l$ are calculated using finite differences.
The segments are moved with the calculated velocity using an automatically selected stiff ODE solver from the \texttt{DifferentialEquations} package \cite{rackauckas2017differentialequations} in Julia language.
$N=80$ segments were used in the calculations. 
The relative error in the equilibrium shape (characterised by the maximum curvature) resulting from this choice was estimated to be $3\cdot10^{-3}$.
During the calculations in this work, the relative length of the filament changed less than $10^{-5}$.

The flow that is produced around the rotating filament is calculated using an integral of near-wall Stokeslets and source doublets distributed along the center-line $\vect y(l)$ of the filament
\begin{equation}
\label{eq:surrounding_velocity}
    u_i(\vect x) = \frac{1}{8\pi\eta} \int_{-L/2}^{L/2} \left( S_{ij}(\vect x,\vect y (l)) + \frac{a^2}{2} D_{ij}(\vect x,\vect y (l)) \right) f^{tot}_j(l) dl,
\end{equation}
where $\vect f^{tot}$ is the total force density (including tension) that is acting on the filament's cross-section. 
For each cylindrical segment it is calculated as $\vect f^{tot} = \zeta_\parallel \vect v^{inext}_\parallel + \zeta_\perp \vect v^{inext}_\perp$, where $\vect v^{inext}$ is the velocity that was obtained after the projection on the space of inextensible motion \cite{nedelec_collective_2007}.

\begin{figure}[h]
\centering
 \includegraphics[width=0.7\textwidth]{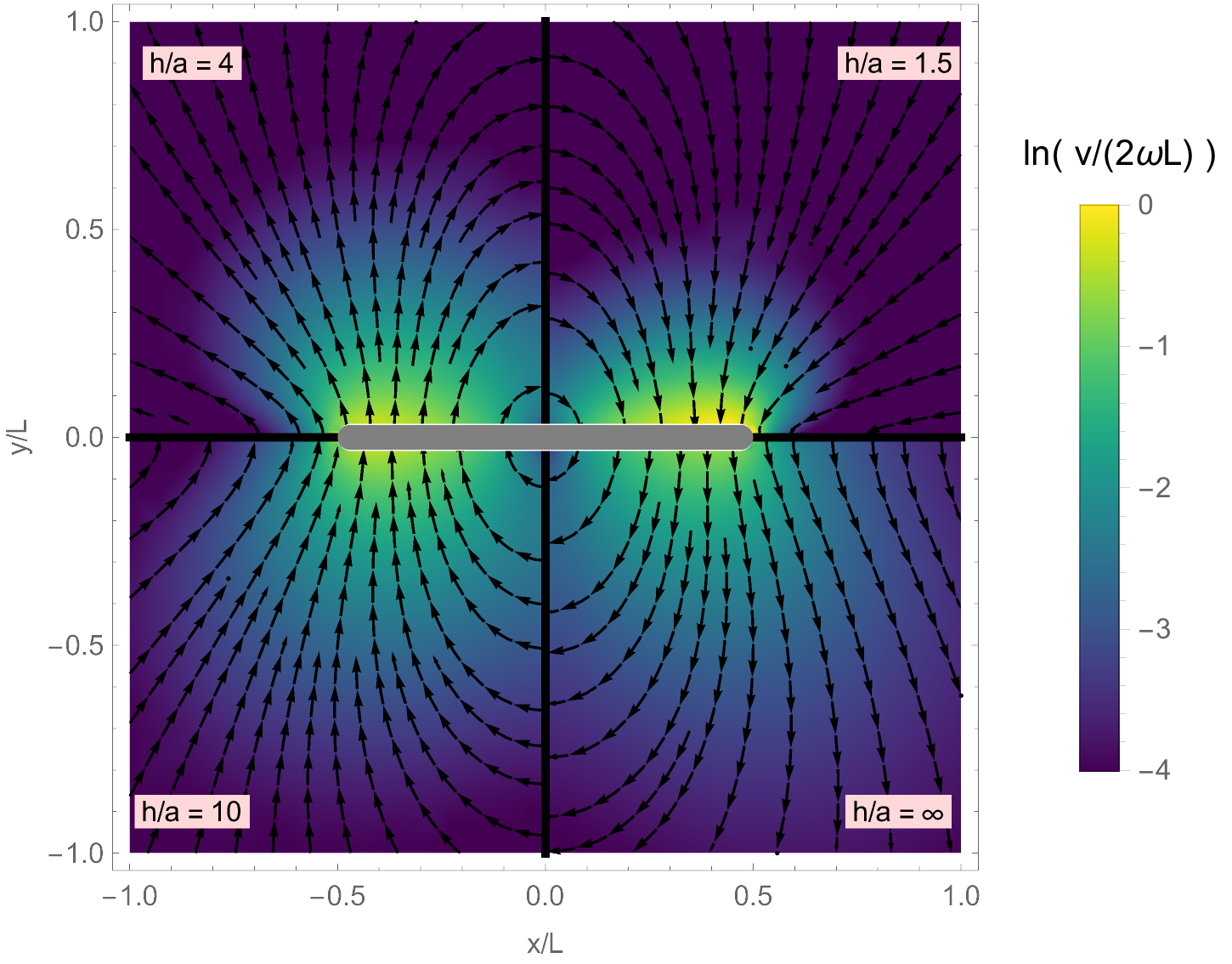}
  \caption{ Velocity field around a straight rod (shown in gray) with length $L$ and radius $a=0.03L$ rotating clockwise with the angular velocity $\omega$. The magnitude $v$ of the velocity is given by the color. Each quadrant corresponds to a different height $h$ above the no-slip surface. In the figure $z=h$.}
  \label{fig:rod_velocity_fields}
\end{figure}

The Stokes flow singularities are given by \cite{blake_fundamental_1974}
\begin{equation}
\begin{split}
    S_{ij} =& \left( \frac{\delta_{ij}}{r} + \frac{r_i r_j}{r^3}  \right) - \left( \frac{\delta_{ij}}{R} + \frac{R_i R_j}{R^3}  \right)  \\
    &+2h(\delta_{j\alpha}\delta_{\alpha k} - \delta_{j3}\delta_{3k})\frac{\partial}{\partial R_k} \left[ \frac{h R_i}{R^3} - \left( \frac{\delta_{i3}}{R} + \frac{R_i R_3}{R^3} \right) \right],
\end{split}
\end{equation}

\begin{equation}
\begin{split}
    D_{ij}f_j=& f_j \left[ \left( \frac{\delta_{ij}}{r^3} - 3 \frac{r_i r_j}{r^5}  \right) - \left( \frac{\delta_{ij}}{R^3} - 3 \frac{R_i R_j}{R^5}  \right)  \right]
    -f_3 \delta_{i \alpha} \frac{6 R_\alpha R_3}{R^5}
    -f_\alpha \delta_{i3} \frac{6 R_\alpha R_3}{R^5} \\
    &+ 2f_3 \delta_{i\alpha} \left( -\frac{3R_\alpha x_3}{R^5} + \frac{15 R_3^2  R_\alpha x_3}{R^7}  \right) 
    +2 f_3\delta_{i3} \left( -\frac{9R_3 x_3}{R^5} + \frac{15 R_3^3 x_3 }{R^7}  \right)\\
    &-2f_\alpha\delta_{i\beta} \left( -\frac{3R_3 \delta_{\alpha\beta} x_3}{R^5} + \frac{15 R_3 R_\alpha R_\beta x_3 }{R^7}  \right) 
    -2f_{\alpha} \delta_{i3} \left( -\frac{3R_\alpha x_3}{R^5} + \frac{15 R_3^2 R_\alpha  x_3 }{R^7}  \right) ,
\end{split}
\end{equation}
where here $\delta_{ij}$ is the Kronecker delta, $\vect r = \vect x - \vect y$, $\vect R = \vect x - \vect y^*$, $\vect x = \{x_1,x_2,x_3\}$, $\vect y = \{y_1,y_2,h\}$, $\vect y^* = \{y_1,y_2,-h\}$. 
The no-slip wall is located at $x_3=0$, and the filament is hovering at a distance $h$ above it.
Repeated Latin indices imply summation $\sum_{j=1}^3$, whereas repeated Greek indices imply summation $\sum_{\alpha=1}^2$.
The Stokeslet term $S_{ij}$ in the velocity integral \eqref{eq:surrounding_velocity} describes the flow that is produced by the force density $\vect f^{tot}$ applied along the center-line of the filament.
Whereas, the addition of the source doublet term $D_{ij}$ ensures that the velocity is constant across a given cross-section of the filament \cite{mori_theoretical_2020}, which gives an improved velocity correction near the filament, but is negligible at large distances.

\section{Velocity field around a straight rotating rod, and determination of height above surface}

It has been reported previously that slender rotating bodies raise vertically above the bottom wall \cite{stikuts_spontaneous_2020, raboisson-michel_kinetics_2020}. 
Accounting for this is important since the no-slip condition on the wall significantly affects the velocity field around the filament (Figure \ref{fig:rod_velocity_fields}). 
Furthermore, height affects the parallel and perpendicular drag coefficients, as shown in eqs. \eqref{eq:drag_coefs}, which in turn impact the calculated shape of the filament.

Assuming that the filament is instead a straight rotating rod, it is possible to calculate the velocity fields around it analytically. 
Let the rod lie momentarily along the $x$ axis $\vect y(l)=\{l,0,h\}$.
The force density exerted by the rod rotating with the angular velocity $\omega$ is
\begin{equation}
    \vect f^{rod} = -\zeta_{\perp} \omega x \vect e_y,
\end{equation}
where $\vect e_y$ is the unit vector along $y$ axis. 
Then the velocity field is given by the integral
\begin{equation}
    u_i^{rod}(\vect x) = -\frac{\zeta_\perp \omega}{8\pi\eta} \int_{-L/2}^{L/2} \left( S_{ij}(\vect x,\vect y (\xi)) + \frac{a^2}{2} D_{ij}(\vect x,\vect y (\xi)) \right)\delta_{j2} \xi d\xi
\end{equation}
where when expanded, the integrand contains terms of the form
\begin{equation}
    \frac{A \xi^n}{(B+(C-\xi)^2)^{m/2}},\quad n=1,2,3 \quad m=3,5,7\quad ,
\end{equation}
which can be readily integrated.
The integration is simple, but cumbersome, therefore it was done using Wolfram Mathematica.
The notebook with the calculated velocity field is given in the supplementary material.

Finally we obtain the expression for the velocity field $\vect u^{rod}(x,y,z,h,\omega,a,L) $, which has been plotted for several values of $h$ in figure \ref{fig:rod_velocity_fields}.
For a straight rod, the velocity field is symmetric with regards to point inversion through the origin. 
Furthermore, with regards to the reflection around either $x$ or $y$ axis, the velocity field is "antisymmetric" - it only acquires a "$-$" sign in front.

\begin{figure}[h]
\centering
 \includegraphics[height=5.99cm]{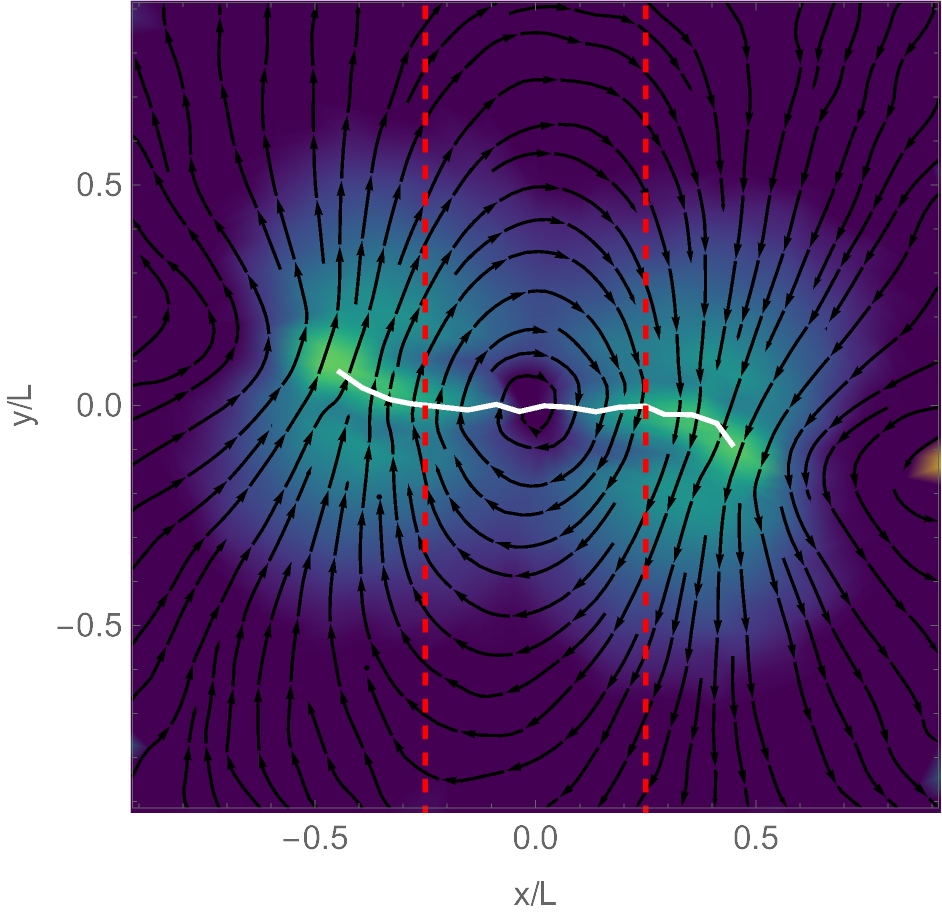}
 \includegraphics[height=6cm]{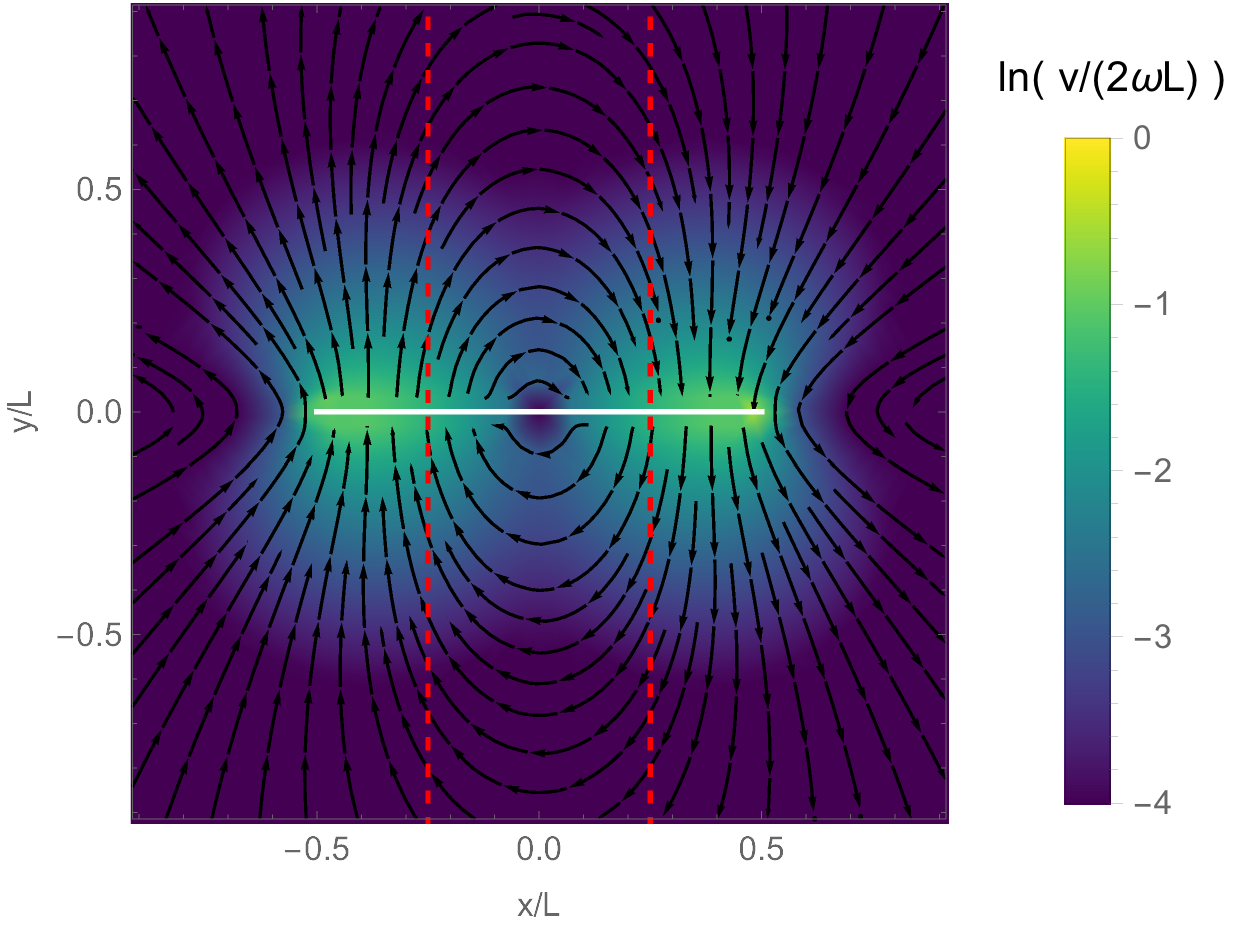}
  \caption{ Illustration of the height fitting. The height $h$ is determined by matching the $y$ component of the velocity field along the red dashed lines located at $x=\pm L/4$. The left side shows the velocity around filament 0 for $f=3$~Hz, the right side shows the best fit (the fitted height is $h\approx0.16 L$) velocity around a straight rod.}
  \label{fig:height_fitting}
\end{figure}

For a flexible ferromagnetic filament the tips bend in the rotation direction, and the velocity field is no longer "antisymmetric" around the $x$ an $y$ axis. 
Nonetheless, we try to determine the height $h$ of the filament by comparing its velocity field to the field produced by a rotating straight rod of the same size.
We plug into $\vect u^{rod}$ the $\omega,a,L$ values from the experiment, and set $z=h$.
To avoid the influence of the bent tips, but to still have a sizable velocity magnitude, we observe the velocity component in the $y$ direction across the filament at $x=\pm L/4$ (figure \ref{fig:height_fitting}). 
We then determine the height $h$ and its error using the \texttt{NonlinearModelFit} function in Wolfram Mathematica.

\begin{figure}[h]
\centering
 \includegraphics[width=0.7\textwidth]{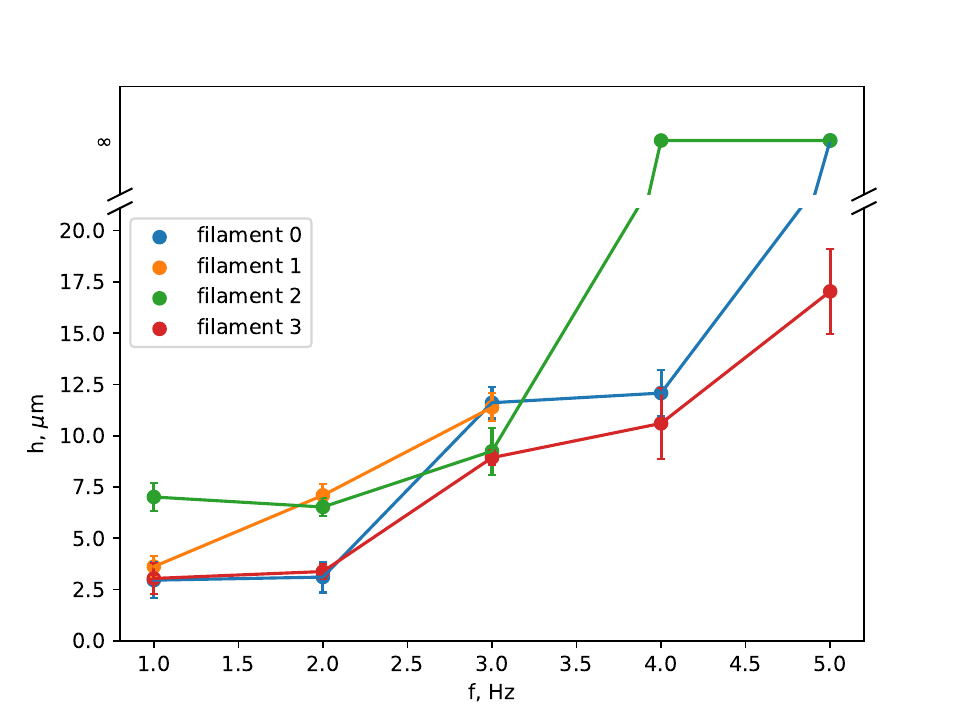}
  \caption{ Fitted heights $h$ at which the filaments rotate depending on the rotation frequency. Infinite height means that the filament is so far from the bottom that the velocity field around it is indistinguishable from that of a filament in an unbounded fluid.}
  \label{fig:heights}
\end{figure}

The fitted heights are shown in figure \ref{fig:heights}. 
There is a tendency that increasing the frequency makes the filament rotate higher up in the sample cell. 
For filament 0 and 2, for large frequencies the velocity field is such that the fitted height $h$ is several meters, and the error of the fit is several times lager than the fitted $h$.
We interpret this result as follows: the velocity field around the filament in these cases is indistinguishable from that of a filament in an unbounded fluid, and this way of determining the height of the rotating filament is no longer accurate.

\section{Comparison of simulated and experimental velocity fields}

Having determined the height $h$ of the rotating filaments, it is possible to calculate their shape and the velocity field around them. 
Other parameters used in the simulations are outlined in table \ref{tab:sim_params}.
The magnetization per unit length $M$ was taken from magnetization measurements \cite{ erglis_experimental_2010, erglis_three_2010}.
The precise experimental determination of the bending modulus $A$ is quite intricate. 
A method that has been used in the past was based on the observation of the rate at which a bent filament straightens once the magnetic field is turned off \cite{zaben_deformation_2020-1}.
This rate, however, is dependent on the height $h$ above the surface. 
The value of $A$ in this work was determined using a height independent method, which will be outlined in more detail in a future paper.
For the viscosity $\eta$ we used the viscosity of water.

\begin{table}[]
    \centering
    \begin{tabular}{c|c|c|c|c|c|c|c}
        filament & beads & $L, \mu m$ & $a, \mu m$ & $\mu_0 H, mT$ & $A, J\cdot m$ & $M, A\cdot m$ & $\eta, Pa\cdot s$ \\
        \hline\hline
         0 & 17 & 72.4 & \multirow{4}*{2.13} & \multirow{4}*{$3.44$} & \multirow{4}*{$2.7 \cdot 10^{-21}$} & \multirow{4}*{$3.3\cdot 10^{-8}$} & \multirow{4}*{$1\cdot 10^{-3}$}\\ 
         1 & 14 & 59.6 & & & & &\\
         2 & 11 & 46.9 & & & & &\\
         3 & 14 & 59.6 & & & & &\\
         \hline
    \end{tabular}
    \caption{Filament parameters used in the simulation.}
    \label{tab:sim_params}
\end{table}

\begin{figure}[h]
\centering
 \includegraphics[width=15cm]{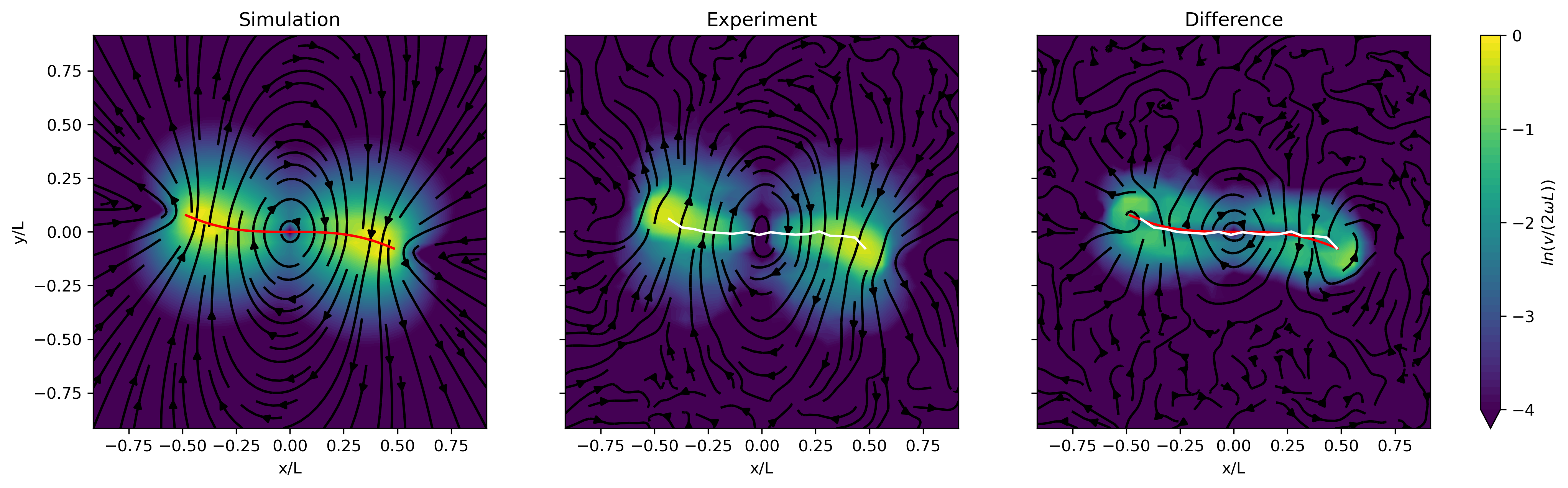}
 \includegraphics[width=15cm]{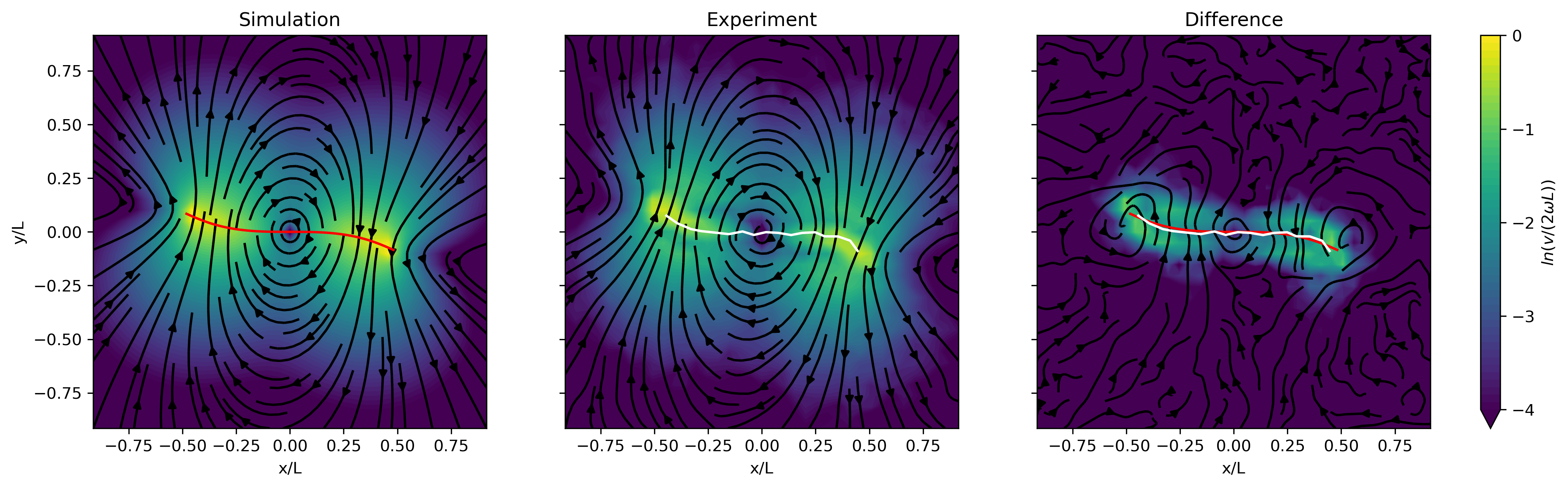}
  \caption{ Velocity field around filament 0 for $f=1$~Hz, $h=2.95~\mu$m (top) and for $f=3$~Hz, $h=11.6~\mu$m (bottom).}
  \label{fig:exp_sim_velocity}
\end{figure}

The calculated velocity field around the filament is sampled at the same points as in the experiment. 
If the sample point is inside the filament, the filament's velocity at that point is used.
A representative comparison between experimental and simulated velocities around the filament is shown in figure \ref{fig:exp_sim_velocity}.
The comparison for other frequencies and filaments is given in the supplementary material.
There is good agreement both for the filament shapes and the distant velocity field. 
The velocity close to the filament is larger in the simulations than in the experiment.
This can be attributed to the experimental underestimate of the velocity next to the filament, as masking step during image processing excludes tracer particle signal close to the filament, where the highest velocities are expected.
Moreover, some of the tracer particles close to the filament move out of the image plane in order to go around it, while PIV can still detect the fluorescence signal.
This results in an undetectable velocity component in $z$ direction and a smaller in-plane velocity.

\section{Characterizing the flow around the filament}

To more quantitatively compare the experimental and simulated velocity fields, we calculate the flux and the circulation of the velocity field $\vect u$ around circular contours centered on the filament's center of rotation. 
In particular, we calculate for different distances $r$ from the origin the integrals 
\begin{equation}
    \oint u_n ds, \quad \oint u_t ds,
\label{eq:circ_integrals}
\end{equation}
where $u_n=\vect u \cdot \vect e_r$ is the component of the velocity field along the radial direction, $u_t=\vect u \cdot \vect e_\phi$ is the component along the azimuthal direction, and $ds$ is the differential of arc length.
$\vect e_r$ and $\vect e_\phi$ are the unit vectors in the radial and azimuthal direction, respectively.
Note that for simulations, to avoid integrating over velocity singularities (eq. \eqref{eq:surrounding_velocity}), when the integral passes over the filament, we instead use the velocity of the filament at that point.

The values of eqs. \eqref{eq:circ_integrals} for filament $0$ are shown in figure \ref{fig:line_integrals}. 
The plots for other filaments can be found in the supplementary material.
The integral of the azimuthal velocity shows a reasonably good quantitative agreement between simulated and measured velocity fields. 
The magnitude of circulation increases as the radius $r$ of the contour increases up until $r\approx L/2$, after which it decays. 
The experimental circulation maximum is systematically more shifted towards larger $r$, which might be explained by the fact that experimentally the beads may be connected with small gaps between them and thus the total length of the filament is larger.

\begin{figure}[h]
\centering
\includegraphics[width=0.49\textwidth]{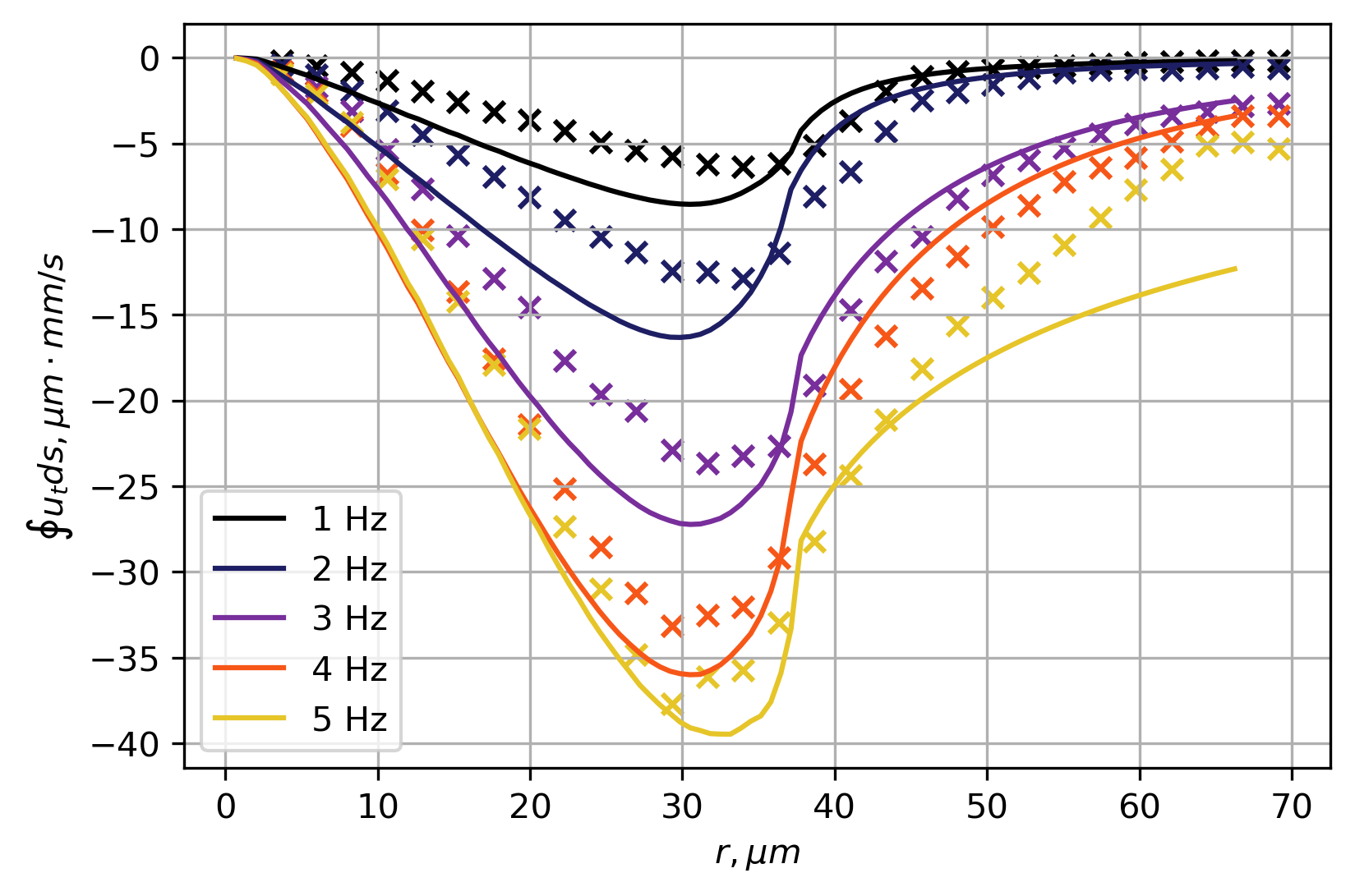}
\includegraphics[width=0.49\textwidth]{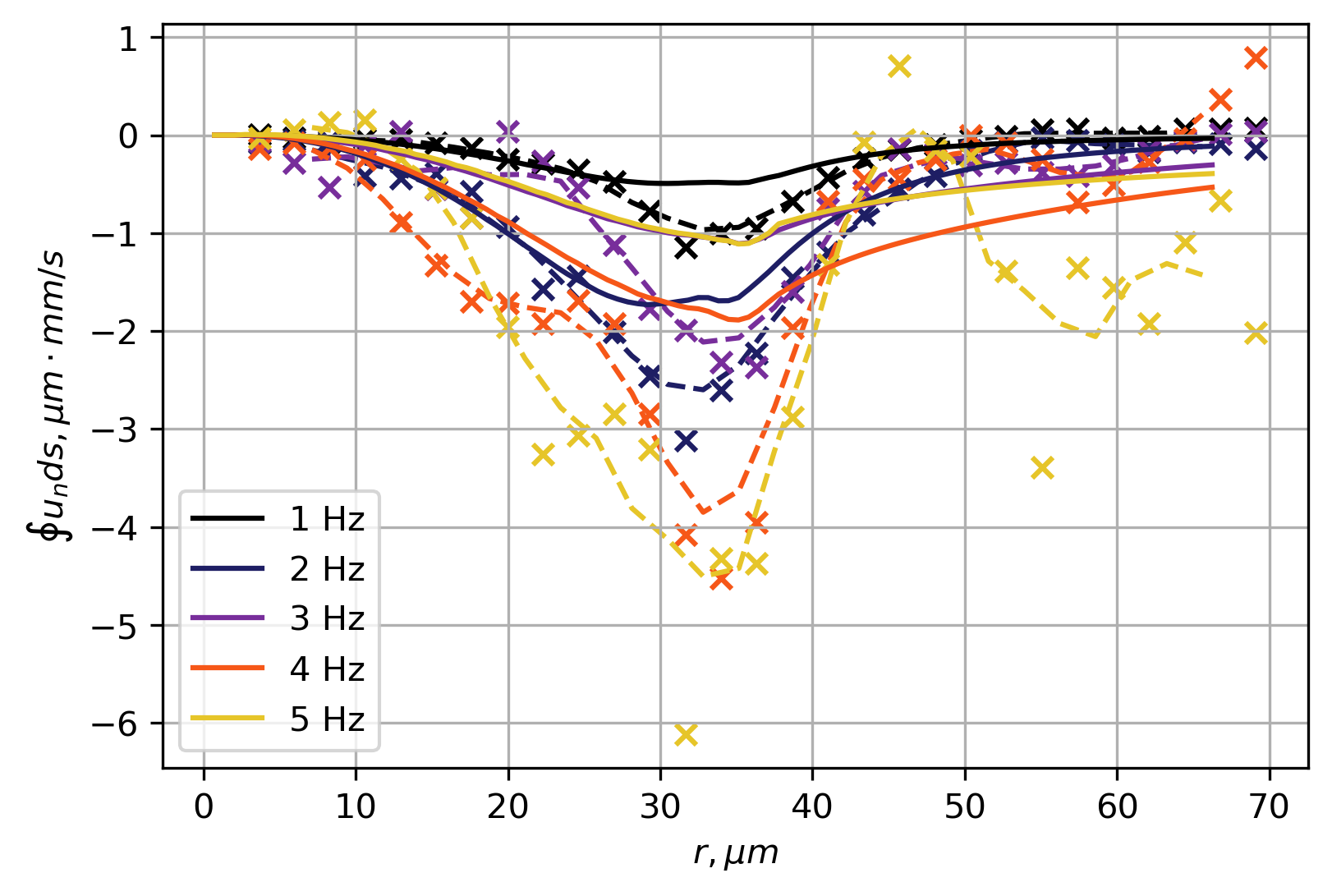}
  \caption{ Line integral of the azimuthal (left) and radial (right) velocity components around circular contours of radius $r$ centered around origin. The solid lines correspond to simulations, $\times$ correspond to experiments and the dashed lines are the experimental moving average to help guide the eye. Filament $0$ data are depicted in these plots.}
  \label{fig:line_integrals}
\end{figure}

A more curious situation can be observed in the graph of the radial velocity integral (right side of figure \ref{fig:line_integrals}). 
There is a velocity flux in the plane of rotation of the filament directed towards the center of rotation.
This flux is most pronounced for the radius of the contour $r\approx L/2$. 
Experimentally the flux is larger, but the qualitative behavior is captured also by the simulations.
Since the surrounding fluid is incompressible this implies that the rotating filament moves it towards the center of the rotation and pushes it out perpendicular to the plane of rotation.

\begin{figure}[h]
\centering
\includegraphics[width=0.4\textwidth]{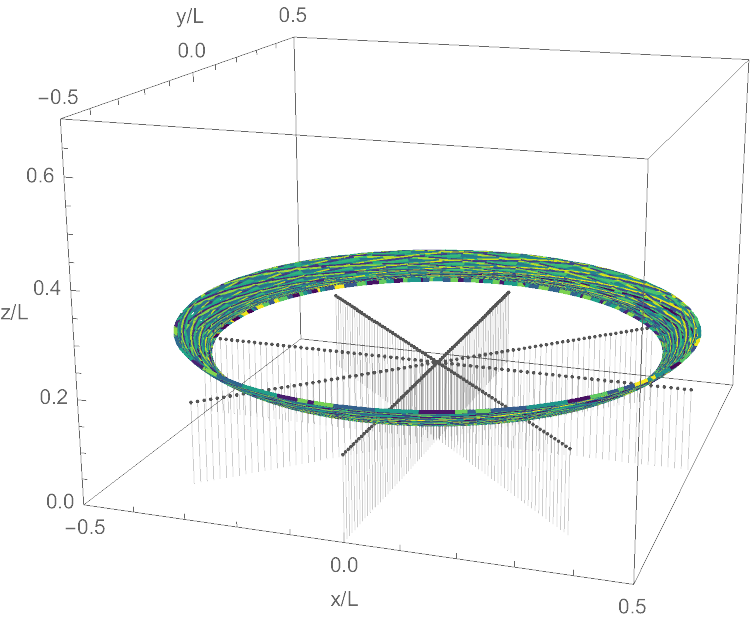}
\includegraphics[width=0.4\textwidth]{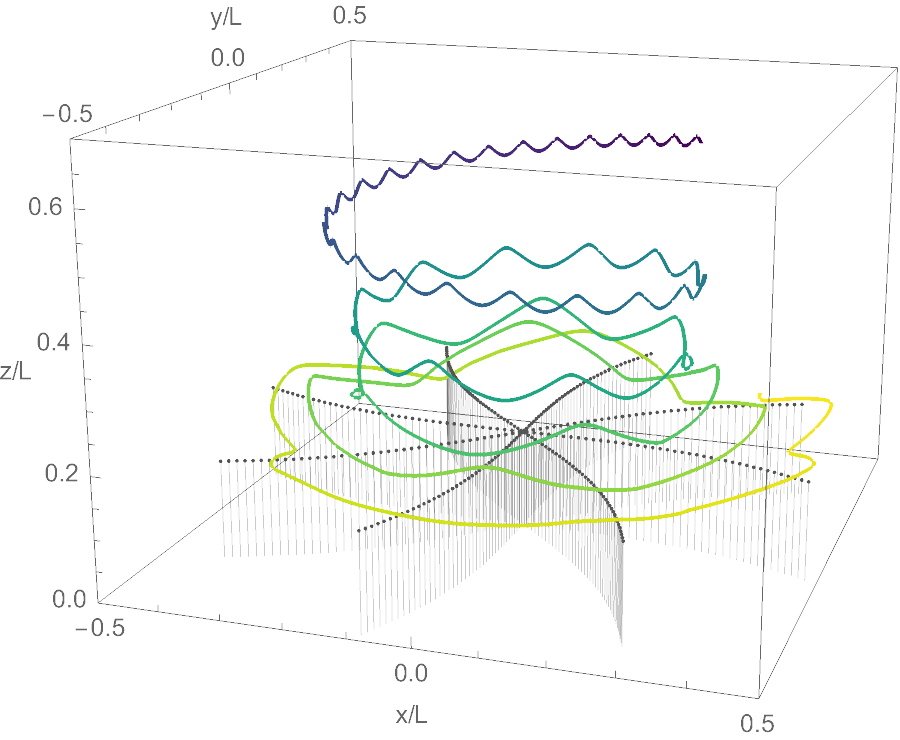}
\includegraphics[width=0.12\textwidth]{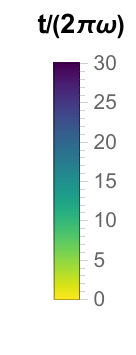}
  \caption{ A sample trajectory around a straight rod (left) and a filament with bent tips (right). Both the rod and the filament undergo 30 rotations clockwise as viewed from above. Time is encoded with color. The no-slip wall is located at $z=0$.}
  \label{fig:trajectories}
\end{figure}

To illustrate this effect more clearly, we have plotted the trajectories of simulated tracer particles around a straight rotating rod and a bent rotating filament (figure \ref{fig:trajectories}). 
Around the rod the trajectory oscillates as the filament passes beneath it, but overall remains in the same height and undergoes roughly a circular motion.
Whereas for a bent filament the particle starts off nearly at the same height as the filament, but over several rotations it gets sucked towards the center and pushed out away from the filament.
It reaches a height of approximately $L/2$ after 30 filament rotations. 

\begin{figure}[h]
\centering
\includegraphics[width=0.4\textwidth]{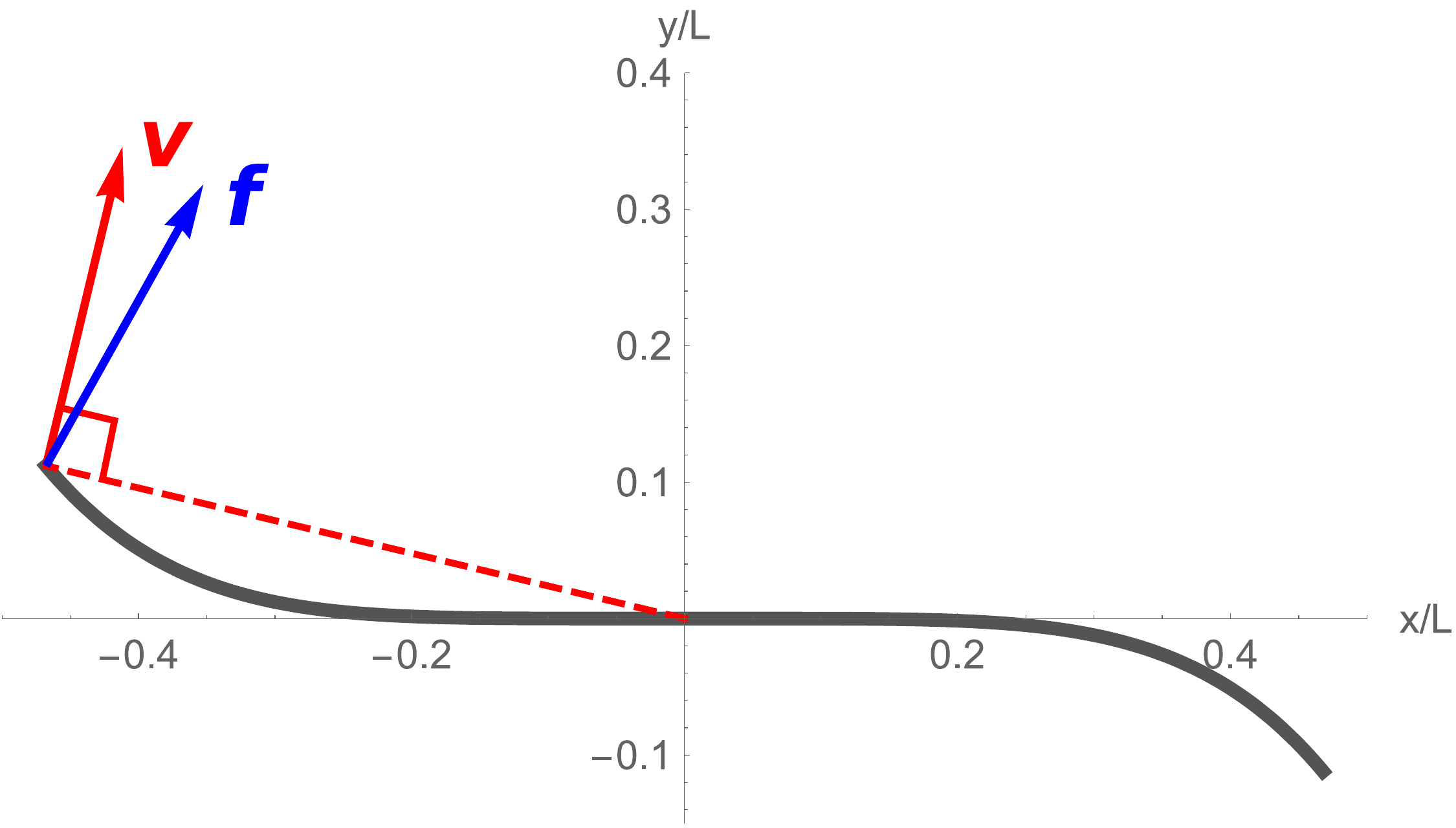}
\includegraphics[width=0.4\textwidth]{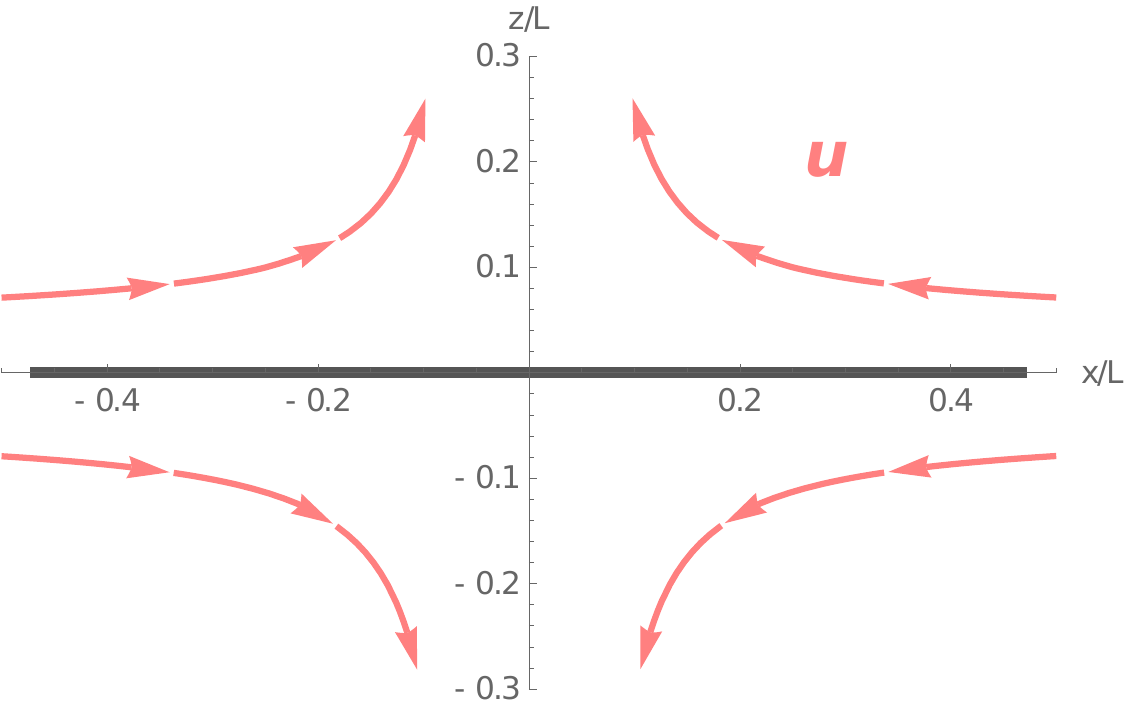}
  \caption{ If the filament has bent tips, there is a force component in the direction of the origin. This results in the background flow $u$ that is directed towards the origin but since the fluid is incompressible, it is ejected along the $z$ axis.}
  \label{fig:schematic}
\end{figure}

Why does a bent rotating filament eject the fluid perpendicular to the plane of rotation and not a straight one? 
We propose the following mechanism (figure \ref{fig:schematic}). 
The velocity $\vect v$ at a particular point on the rotating filament is perpendicular to the radius vector from the center of rotation.
However, due to the anisotropy of drag $\zeta_\parallel < \zeta_\perp$, the force density $\vect f$ that the filament exerts on the fluid at this point can have a component along the radius vector.
In particular if the tips are bent in the direction of rotation (as it is the case for magnetic filaments), there is a force component towards the center of rotation. 
This forces the fluid towards the center and due to incompressibility it is ejected perpendicular to the plane of rotation.
In an infinite fluid it is ejected both up and down by the same amount, however, if the filament is rotating near a wall, fluid is mostly ejected in the direction away from it.
This effect has implications for microfluidic mixing, where we propose that, all else being equal, magnetic filaments will outperform straight rigid rod mixers by transporting the fluid also perpendicular to the rotation plane.

\section{Conclusions}

In summary, we have shown that micro-particle image velocimetry can be successfully used to measure the flow around microscopic filaments.
However, to increase the precision of the method, averaging of the velocity field is needed, which can be readily done for symmetric, time-periodic motion such as rotating filaments.

Our measurements indicate that rotating filaments hover above the bottom and the hovering height increases with the increase of the rotation frequency.
This height greatly influences the velocity field around the filament.
Fitting the numerically calculated velocity field of a straight rotating rod to the experimentally measured velocity fields allow to determine this height.
From physics point of view, the height above the bottom influences the drag force that the filament experiences.
This in turn has an effect on the filament's shape.
Decreasing the height increases the drag, which increases the bending of the tips. 
This has the same effect as increasing the frequency of the field.
If the height above the bottom is taken into the account, the elastic rod model can reasonably accurately calculate both the shapes and the velocity fields around rotating ferromagnetic filaments.

Our experimental and numerical results show a good agreement.
By this we discover that rotating flexible ferromagnetic filaments, as opposed to rigid straight rods, suck in the fluid in the rotation plane and push it out along the axis of rotation.
This effect is the result of the bent tips of the filament.
The induced 3D flow gives an additional applicability in micro-mixing.

\section*{Data availability statement}
Data will be made available in Zenodo repository, following the OpenData principles, as a part of community "Magnetics and Microhydrodynamics: From guided transport to delivery". 

\section*{Acknowledgment}
A.P.S., I.D., M.\v{S}. and G.K. acknowledge the funding by the Latvian Council of Science, project A4Mswim, project No. lzp-2021/1-0470. A.Z. acknowledges the supprot from the European Union's Horizon 2020 research and innovation programme project MAMI under grant agreement No.766007. A.C. \& R.L. acknowledge the funding by the Latvian Council of Science, project BIMs, project No. lzp-2020/1-0149 and the financial support from the M-era.net project FMF No.1.1.1.5./ERANET/18/04.

\printbibliography

\end{document}